\documentstyle[emulateapj,onecolfloatx]{article}

\newcommand{\beq}{\begin{equation}}
\newcommand{\eeq}{\end{equation}}

\def\beqa{\begin{eqnarray}}
\def\eeqa{\end{eqnarray}}
\newcommand{\lsim}{\lesssim}
\newcommand{\gsim}{\gtrsim}
\def\p{\partial}

\def\xv{{\bf x}}
\def\vv{{\bf v}}

\lefthead{Peebles, Phelps, Shaya \&\ Tully}
\righthead{Transverse Velocities of Galaxies}

\begin{document}

\twocolumn[

\submitted{}

\title{Radial and Transverse Velocities of Nearby Galaxies}

\author{P. J. E. Peebles$^{1}$, S. D. Phelps$^{1}$, 
Edward. J. Shaya$^{2,3}$, and  R. Brent Tully$^{4}$}
\affil{{}$^1$Joseph Henry Laboratories, Princeton University,
Princeton, NJ 08544\\
{}$^2$ Raytheon ITSS, Goddard SFC, Greenbelt MD 20771\\
{}$^3$  U. of Maryland, Physics Dept., College Park MD, 20743\\
{}$^4$Institute for Astronomy, University of Hawaii,
Honolulu, HI 96822\\}

\begin{abstract}
Analysis of the peculiar velocities of galaxies should
take account of the uncertainties in both redshifts and
distances. We show how this can be done by a numerical
application of the action principle.  The method is applied
to an improved catalog of the galaxies and tight systems of
galaxies within 4$h_{75}^{-1}$~Mpc, supplemented with a coarser
sample of the major concentrations at 4$h_{75}^{-1}$~Mpc to
20$h_{75}^{-1}$~Mpc distance. Inclusion of this outer zone
improves the fit of the mass tracers in the inner zone to
their measured redshifts and distances, yielding best fits
with reduced $\chi ^2$ in redshift and distance in the range
1.5 to 2.  These solutions are based on the assumption that
the galaxies in and near the Local Group trace the mass,
and a powerful test would be provided by observations of
proper motions of the nearby galaxies.  Predicted transverse
galactocentric velocities of some of the nearby galaxies are
confined to rather narrow ranges of values, and are on the
order of 100 km~s$^{-1}$, large enough to be detected and
tested by the proposed SIM and GAIA satellite missions.

\end{abstract}  

\vspace{3 mm}\

]

\section{Introduction}

This paper continues the analysis of the dynamics of relative
motions of the galaxies in and near the Local Group. 
In previous numerical applications of the action principle
either the present distances of the assumed mass tracers are
given, and the measure of goodness of fit is the difference
between model and catalog redshifts (\cite{Peeb89} 1989),
or the present redshifts are given, and the measure of
goodness of fit is the difference between model and catalog
distances.\footnote{The use of variants of the action
principle to find solutions to the equation of motion with
given present redshifts rather than distances is discussed
by \cite{Maro} (1993), \cite{Peeb94} (1994), and \cite{SPT}
(1995). We use a canonical transformation of the action, which
was introduced, we believe independently, in \cite{Schmoldt}
(1998), \cite{Whiting} (2000), and \cite{SDP} (2000).} Since errors
in redshift and distance both are important we have adapted the
numerical action method to allow relaxation to a stationary
point of the action and a minimum of the sum of the squares of
the weighted differences of model and catalog redshifts and
distances. The method is described in \S~3. 

In \S~4 we present an application of the method to the catalog
discussed in \S~2. Recent advances in measurements of distances
of nearby galaxies are incorporated in a catalog of the
distances and radial velocities of the galaxies and tight
concentrations of galaxies within 4$h_{75}^{-1}$~Mpc.
(Hubble's constant is $H_o=75h_{75}$ km~s$^{-1}$~Mpc$^{-1}$).
This inner zone catalog is supplemented by a coarser catalog of
distances  
and radial velocities of the main galaxy concentrations in an
outer zone at 4--20$h_{75}^{-1}$~Mpc distance. We work with the
assumption that the luminosities of the galaxies trace the mass.
Two control cases support this assumption: (i) inclusion of
the outer 
zone in the dynamical solutions improves the fit to the measured
redshifts and distances in the inner zone, and (ii) replacing the
angular positions by random positions on the sky, while keeping
the catalog redshifts and distances, worsens the fit.

The application presented here is meant to illustrate the new
numerical method and catalogs; we hope to present a systematic
study of parameters in a later paper. With the parameter choices
for cosmology and mass-to-light ratios ($M/L$, using blue-band light)
listed in \S~4 we get
values of the reduced $\chi^2$ for distances and redshifts that
in the best cases are between 1.5 and 2, close enough to unity to
add to the evidence that our model is a useful 
approximation to reality. A considerably stronger test, from
measurements of transverse motions of nearby galaxies, is
discussed in \S~5. 

\section{Catalogs of Nearby Galaxies}

Our compilation of mass tracers and potential targets
for measurements of proper motions is restricted by three
considerations. (1) Accurate distances are needed to constrain 
models. (2) In this first exploration of the method we want to 
restrict the number of orbits to reconstruct. (3) Complex --- 
strongly nonlinear --- orbits can only be recovered with an
extensive exploration of initial conditions. The first two
led us to study about 40 objects. The third compels the merging
of some closely adjacent galaxies into single entries in our
catalog.

Regarding the last point, it is possible in principle to
find numerical action solutions in cases where galaxies
have already made close passages, as long as there has
not been significant exchange of orbital and internal energy.
For example, the interactions between  M81, M82, and NGC 3077 left
extended tidal streams of neutral hydrogen that provide
detailed constraints on the relative motions (\cite{M81HI}
1994), a case that would be fascinating to study. 
However, the current
work has a much less ambitious goal. If galaxies are so near
to each other that the orbits may be complex, we merge the
information on these systems: the catalog entries are the sums
of the luminosities and the luminosity-weighted positions and
velocities.  What is meant by `near to each other' depends on
the available information. For galaxies that are close to us
we have better relative discrimination of positions and a more
complete census of the major mass influences. Consequently,
we attempt to explore somewhat more complex situations in the
nearest volume.

Our catalog has `inner' and `outer' zones. The former,
within 4$h_{75}^{-1}$~Mpc distance, is the main sample in
this study. About $90\%$ of the light in the inner zone resides
in just 8 of the objects. If mass and light are strongly correlated
then most of the inner catalog mass is associated with
these few giant systems.  It
is important for the dynamical analysis that these few big
objects be located as precisely as possible in position and
velocity. The many small galaxies
make a minor contribution to the light and, we are
assuming, to the mass. We 
seek to include as many of these objects as feasible because
they are valuable probes of the gravitational potential,
but their exclusion would have little effect on the
dynamics. Thus we include all the massive galaxies in the inner
zone, whatever the reliability of their distance estimates,
but only dwarfs that are sufficiently isolated to have simple
orbits and for which we have good
distance estimates.

Thanks especially to the Hubble Space Telescope, reasonably good
distances are emerging for many galaxies in and at the fringe
of the Local Group. Two techniques are providing most of the
data: the correlation between pulsation period and luminosity
of Cepheid variable stars (eg., \cite{Freedman} 1994) and
the calibrated luminosity of the tip of the Red Giant Branch for
evolved low-metallicity stars (eg., \cite{Cole} 1999).  Each
can provide distances with 10\%\ standard deviation.  There is
a further 10\% uncertainty in the underlying zero points, 
but this normalization is close to a common factor in all distances
and so enters only as a scale factor in the dynamics. 
The distance scale used in this paper is consistent with 
H$_{\circ}=75$~km~s$^{-1}$~Mpc$^{-1}$ (\cite{TP00} 2000;
\cite{Tonry} 2000); a zero point shift is accommodated by the
factor $h_{75}={\rm H}_{\circ}/75$.
Distance
measurement coverage of individual galaxies falls rapidly 
beyond 4$h_{75}^{-1}$~Mpc.

Some comments on the characteristics of the nearby structure
are in order. Almost all galaxies within 1$h_{75}^{-1}$~Mpc
have negative systemic velocities, which suggests this region
is gravitationally bound but not yet relaxed.  The galaxies
in this region make up the Local Group. This condensation is
part of an expanding filamentary structure, with the nearest
big galaxies close to each other in the sky in what are called
the Maffei--IC342 and M81 groups. The filament originates in
the northern Galactic hemisphere at a conjunction of filaments
near the Ursa Major and Coma I clusters, passes through small
knots of galaxies in the constellations of Canes Venatici and
Ursa Major, through our position, and can be followed in the
southern Galactic hemisphere out to the NGC 1023 Group. The
thread of galaxies that makes up what has been called the
Sculptor Group can be viewed as a minor fragment of the main
filament. It seems this filament is paralleled by a second
one that originates in the Galactic north at the Virgo Cluster,
passes through a big knot including the Sombrero galaxy,
then passes nearest to us at the Centaurus Group, through
the Galactic plane in the vicinity of Circinus galaxy, and
onward to meet with the Telescopium-Grus Cloud. \cite{TF87} 
(1987) provide maps of these structures.

The outline of these features can be seen in Figure~1, though
in skeleton form because we have plotted only the important
dynamical constituents and those that serve as useful
test particles. Our inner zone catalog in Table~1 is shown as the
filled symbols. This sample is thought to be complete in
luminous objects ($L_B>10^{10}~L_{\odot}$) within 4$h_{75}^{-1}$~Mpc, 
and we are assuming
it is a good sample of the mass.
The unnamed objects are dwarfs with decent distance
measurements. All the inner zone objects are concentrated
near the equatorial plane in supergalactic coordinates.
If light traces mass the major gravitational influence on the
Local Group from the inner zone is the Maffei/IC342/M81 complex, 
with some influence from the Centaurus Group, a
minor effect from the modest galaxies in the Sculptor Group,
and little else.  There are no important mass elements near
the boundary of the inner zone. This degree of isolation and
the sharp decline in the completeness
of distance measurements led us to place the boundary for the
main inner zone at 4$h_{75}^{-1}$~Mpc.

\begin{figure}[ht]
\plotfiddle{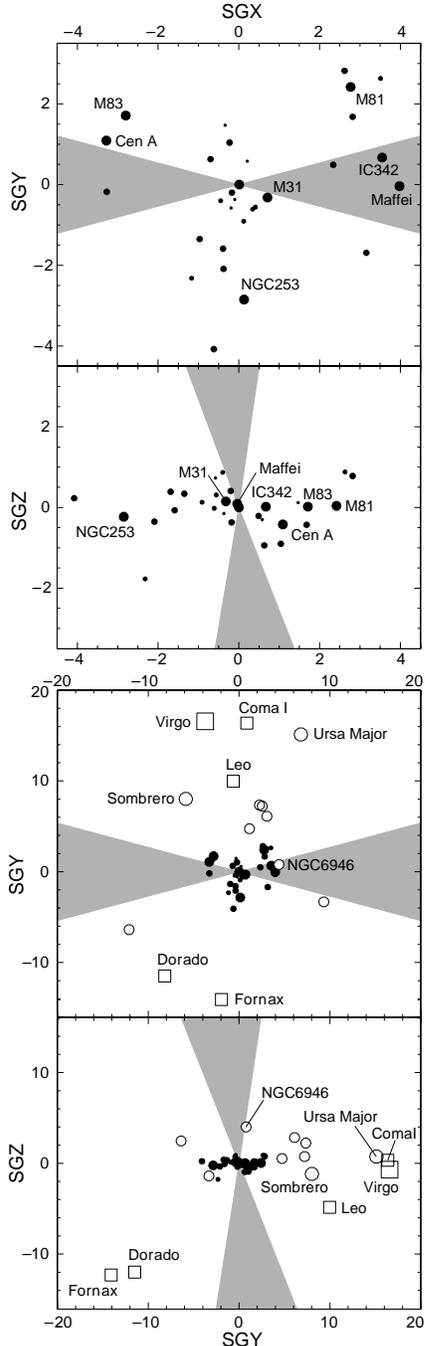}{6.8in}{0}{86}{86}{-260}{-95}
\caption{
Projected locations of catalog objects.  
{\it Top pair of panels:} Region within $\pm 4 h_{75}^{-1}$~Mpc
from two orthogonal viewpoints in supergalactic coordinates.
{\it Bottom pair of panels:} Region within $\pm 20 h_{75}^{-1}$~Mpc
from the same orthogonal viewpoints.  
Shaded wedges identify zones of Galactic obscuration.
}
\end{figure}

Gravitational interactions with more
distant mass concentrations affect the dynamics within our
inner zone. The nearest substantial exterior concentrations
of galaxies are at distances of 15--20$h_{75}^{-1}$~Mpc.  Then at
30--60$h_{75}^{-1}$~Mpc there are still larger concentrations
in the Norma-Hydra-Centaurus and Perseus-Pisces supercluster
regions.  For this paper we ignore the big structures beyond
$30h_{75}^{-1}$~Mpc but try to account for the significant
structures within an outer zone at 4--$20h_{75}^{-1}$~Mpc
distance. The distribution of important galaxy concentrations in
this outer zone is simple enough to be modeled by the 14 objects
in Table~2. These groups or clusters (numerical names from \cite{T88}
1988) are shown as open symbols in 
Figure~1, and the eight that cause the largest gravitational tides
at our position (based on  mass divided by distance cubed) 
are identified by name
in the bottom pair of panels in the figure. 

The Virgo Cluster is not an outstandingly dominant contributor  
to the local luminosity, but there is good evidence that it is 
a larger contributor to the mass. Numerical action modeling,
which mostly refers to the low density environments where most 
galaxies are found, indicates the cosmological mass density
parameter is $\Omega_{m} = 0.25 \pm 0.2$ (95\% formal probable
error) with mass-to-light ratio 
$M/L \sim 150h_{75}~M_{\odot}/L_{\odot}$ in typical galaxy  
environments (\cite{SPT} 1995; \cite{Tully} 1999). The evidence
that there is much more mass in the Virgo Cluster than would
be inferred from this value of $M/L$ includes the velocity
dispersions of the early and late-type galaxies near the core,
the high velocities of galaxies infalling on first approach
to the cluster (\cite{TS84} 1984; 1998), and the
virgocentric flow derived from the distances in the surface
brightness fluctuations survey (\cite{Tonry} 2000).
\cite{TS98} (1998) conjecture that the Virgo Cluster is not
the only nearby place where $M/L$ is large, that regions with
short dynamical times ($t_{collapse} \sim 10^9$ years) have
high $M/L$, and that such regions are readily identified as
nests of elliptical and S0 galaxies.  It is argued that these
regions have $M/L$ at least five times that of regions with
long dynamical times ($t_{collapse} \sim 10^{10}$ years). If
so, four other systems with large $M/L$ within 
20$h_{75}^{-1}$~Mpc are the E/S0 knots associated with the
Fornax, Dorado, Coma I, and Leo 
clusters/groups.  These entities are located by large boxes
in the bottom two panels of Figure~1; we assign them 
$M/L$ five times that of the nine low density groups with long
dynamical times and mainly late-type galaxies in the outer zone.  
The most massive of the low $M/L$
members of the outer zone, the Sombrero and Ursa Major
clusters, are identified by the larger open circles in Figure~1.  
The NGC 6946 group is the nearest of the outer zone systems at
6$h_{75}^{-1}$~Mpc (a distance that still is quite uncertain).

The mass we assign to the
Virgo Cluster is four times that of the second most massive
system in Table~2, and larger than the sum of all the other
masses within 20$h_{75}^{-1}$~Mpc.  A simple measure of
the interactions among inner and outer zone objects is
given in the last
column in Table~2: the relative contribution of each outer
object to the present tidal field at our position. One sees again
the importance of the Virgo cluster, and the not insignificant
sum of contributions from the rest of the outer zone objects.

The luminosities of all the galaxies in the outer zone that are
{\it not} in the 14 objects in Table~2 add up to a luminosity
density that is comparable to the cosmic mean used by \cite{SPT}
(1995). That is, the 14 mass tracers in the outer zone approximate
the {\it overdensity} of luminosity, and plausibly of mass, within
20$h_{75}^{-1}$~Mpc.

Our numerical analysis assumes distance measurement errors
are symmetrically distributed in the distance moduli, with the
standard deviations $\sigma _\mu$ listed in Tables~1 and~2. The
standard deviations $\sigma _{cz}$ in the redshifts in Table~1
are measurement errors, while the $\sigma _{cz}$ listed in Table~2
are mean deviations of the velocity dispersions within these 
clusters and groups of galaxies. A crude model for the motion of
each galaxy relative to the dark halo it is supposed to trace is 
presented in \S~4. 

The objects near the outer boundary of the outer zone likely are
strongly influenced by structures beyond 20$h_{75}^{-1}$~Mpc that
are not in our analysis. Hence although orbit computations are
{\it constrained} by the distances and velocities, with their 
uncertainties, in both zones, the solutions are {\it
evaluated} only from the fits to the redshifts and distances in
the inner zone. 

\section{Solutions that Minimize $\chi ^2$}

\subsection{Numerical Action Method}

This initial discussion,  for simplicity, deals
with a single particle. A solution to the equation of motion
generally has neighboring solutions with neighboring values
of the present distance and  redshift, and we seek the member
of this family that minimizes
\beq
\chi ^2 = (z - z_c)^2/\sigma _z^2 + 
(\mu - \mu _c)^2/\sigma _\mu ^2,
\label{eq:chis}
\eeq
where the redshift and distance modulus in the solution are $z$
and $\mu$, the measured catalog values are $z_c$ and $\mu _c$,
and their 
standard deviations are $\sigma _z$ and $\sigma _\mu$. The
relaxation of a single-particle orbit to stationary points
of the action and of $\chi ^2$ is applied separately to each
member of the catalog and iterated until the gradients of the
action and $\chi ^2$ vanish to machine accuracy. Relaxation
to a minimum of the sum of the $\chi ^2$ over all particles
(except the reference), rather than iterated relaxation to
minima of single-particle $\chi ^2$, is feasible but the matrix
inversion is a considerably heavier computation.

We represent the orbit of a particle by its comoving positions
$\xv _n$ at a sequence of time steps $t_n$, $1\leq n\leq n_x$,
together with the present distance $d$ at time $t_o=t_{n_x+1}$
and given present angular position, as in \cite{Peeb95}
(1995). The action function is
\beq
A = S + H_od^2/2 - d(cz + \vv _{\rm MW}\cdot\hat x),
\label{eq:A}
\eeq
where 
\beqa
S &=& \sum _{n=1}^{n_x}{a_{n + 1/2}^2\over 2}
{(\xv _{n+1}-\xv _n)^2\over (t_{n+1} - t_n)} \nonumber\\
&-& (t_o - t_{n_x + 1/2})V_o \nonumber\\
&-& \sum _{n=1}^{n_x}(t_{n+1/2} - t_{n-1/2})V_n.
\eeqa
The term $S$ approximates the time integral of
the kinetic minus
the potential energies. The potential for particle $i$ is 
\beq
V = -G\sum _{j\not= i}{m_j\over a(t)x_{ji}}
- {\Omega H_o^2a_o^3\xv _i^2\over 4a}. 
\label{eq:V}
\eeq
The physical distance between particles $i$ and $j$ is
$a(t)x_{ij}$. The second term in $V$ refers the potential
to the background cosmological model with present Hubble
constant $H_o$ and density parameter $\Omega$ in matter
with mass density that varies inversely as the cube of
the expansion parameter $a(t)$. The terms added to $S$ in
equation~(\ref{eq:A}) represent a canonical transformation
that in effect exchanges radial positions and momenta, thereby
changing the boundary condition from fixed present distance
to fixed present radial velocity (the details of which are
presented in \cite{SDP} 2000). The present radial velocity of
the particle relative to the Milky Way is $cz$, and the last
term in equation~(\ref{eq:A}) adds the component of the peculiar
velocity $\vv _{\rm MW}$ of the Milky Way in the direction $\hat
x$ of the particle. (The velocities are such that the center
of mass moves according to eq. [7] in \cite{Peeb94} 1994). The
derivative of the action with respect to the present distance is
\beqa
{\p A\over\p d} &=& {a_{n_x+1/2}^2\over a_o}
{\hat x\cdot (\xv _o-\xv _{n_x})\over t_o-t_{n_x} }\nonumber\\
&-& (t_o-t_{n_x + 1/2}){\p V_o\over\p d} \nonumber\\
&+& H_od - cz - \vv _{\rm MW}\cdot\hat x.
\label{eq:dAdd}
\eeqa
The first term on the right hand side is the radial canonical
momentum of the particle a half time step before the
present, where $\hat x$ is the unit vector to the present
particle position. The second term brings the radial momentum
to the present epoch. When $\p A/\p d=0$ this agrees with the
radial peculiar velocity given by the last three terms. One
similarly sees that the conditions $\p A/\p\xv _n =0$ represent
the equation of motion in leapfrog approximation (\cite{Peeb95}
1995). The motion of the Milky Way is at a stationary point
of the function $S$ of the coordinates of this orbit, with
the boundary condition that the Milky Way ends up at the origin.

To simplify discussion of the relaxation to stationary points
of $A$ and $\chi ^2$ let $x_\alpha$ represent the $3n_x+1$
coordinates for the orbit of a particle (or the $3n_x$
coordinates for the Milky Way), with $\alpha =0$ for the
present distance $x_0=d$. The first and second derivatives
of the action with respect to these variables are $A_\alpha$
and $A_{\alpha\beta}$. The square of the gradient of the
action for particle $i$ is $\Gamma _i= \sum (A_\alpha )^2$. The
coordinate shift
\beq
\delta x_\alpha = -\epsilon \sum_{\beta}A^{-1}{}_{\alpha\beta }A_\beta 
\label{eq:deltax}
\eeq
reduces $\Gamma _i$ if $\epsilon$ is small enough. If
$x_\alpha$ is close to a stationary point, so $A$ is close
to a quadratic function of the distance from the stationary
point, then $\epsilon =1$ brings the orbit much closer to the
stationary point, whether a local extremum or saddle point.

Now consider neighboring stationary points of the action, 
where $A_\alpha =0$, belonging to neighboring redshifts 
$z$ and $z+dz$. The result of differentiating $A_\alpha =0$ with
respect to $z$, and remembering that the coordinates $x_\alpha$
are functions of redshift and that $z$ also appears explicitly in 
$A_0$ (eq.~[\ref{eq:dAdd}]) is
\beq
A_{\alpha\beta }{\p x_\beta\over\p z} = c\delta _{\alpha,0}.
\label{eq:dxdz}
\eeq
We see that in a family of solutions related by  
continuously varying the redshift $z$ the derivative of the
present distance $d$ with respect to the redshift is
\beq
{\p d\over\p z} = cA^{-1}{}_{00}.
\label{eq:dddz}
\eeq
The inverse of the matrix of second derivatives of $A$
thus shows how to move the orbit toward a stationary point
of $A$ (eq.~[\ref{eq:deltax}]) and how the present distance
at a stationary point of $A$ changes when the redshift is
adjusted. We use the latter to relax to a minimum of
$\chi ^2$. The derivative of $\chi ^2$ (eq.~[\ref{eq:chis}])
with respect to the redshift satisfies
\beq
\sigma _z^2{\p\chi ^2\over\p z}/2 = z - z_c + R {\p d\over\p z}
{\ln d/d_c\over d},
\eeq
where
\beq
R = \left( {5\over 2.303}{\sigma _z\over\sigma _\mu }\right) ^2.
\eeq
We find that it is a good numerical approximation to suppose 
$\p d/\p z$ is independent of $z$. In this case the second
derivative satisfies 
\beq
\sigma _z^2{\p ^2\chi ^2\over\p z^2}/2 = 
1+{R\over d^2}\left(\p d\over\p z\right) ^2 (1 -  \ln d/d_c).
\label{eq:chipp}
\eeq
The analog of equation~(\ref{eq:deltax}) for a redshift
adjustment that moves the orbit toward the minimum of $\chi
^2$ is
\beq
\delta z = -\epsilon {\p\chi ^2/\p z\over\p ^2\chi ^2/\p z^2}.
\label{eq:deltaz}
\eeq

To sample the families of solutions we start from random
orbits with the positions $\xv _n$ at $1\leq n\leq n_x$ placed
independently at random in the sphere that contains the mass of
the system of objects in the background cosmological model.
The iterative relaxation first uses the matrix inverse
$A^{-1}{}_{\alpha\beta}$ to find $\p d/\p z$ and hence the
redshift adjustment $\delta z$ (and the accompanying adjustment
to $\p A/\p d$) that moves the orbit toward a minimum of 
$\chi ^2$, and then uses the same matrix inverse to find the
coordinate adjustments $\delta x_\alpha$ that move the orbit toward a
stationary point of $A$. 

\subsection{Numerical Fixes}

We expect colleagues who wish to explore applications of this
method would do well to seek independent ways to address the
numerical problems arising, so as to improve our inelegant and
likely inefficient fixes, but to aid reproducing our results we
list the main elements of our procedure.

The square of the gradient of the action summed over all
objects $i$ and position coordinates $\alpha$ is
\beq
\Gamma = \sum _{i}\Gamma _i = \sum _{i,\alpha}(\p A/\p
x_{i,\alpha})^2.
\label{SOS}
\eeq
With the velocity unit 100~km~s$^{-1}$, length unit 1~Mpc,
and mass unit $1\times 10^{11}$ solar masses, the square of
the gradient of the action for the initial random orbits
typically is $\Gamma\sim 10^8$. At $\Gamma\lsim 10^{-10}$
further iterations usually reduce the square of the gradient of
$A$ to zero to machine accuracy (32 bit) without appreciable
perturbations to the orbits. In the solutions presented here
the iteration stops at $\Gamma\sim 10^{-20}$, at which point
$(\p\chi ^2/\p z)^2$ is similarly small.
The iteration can approach a point where $\Gamma$ has a local
minimum but not a zero. In our experience when this happens
$\Gamma$ is dominated by the contribution from one object,
so when $\Gamma _i>\Gamma /2$ for object $i$ its orbit alone
is adjusted until $\Gamma _i < 0.01\Gamma$ or to a maximum
of 300 iterations. If this happens for particle $i$ more than 100
times we choose a new random orbit for the object and relax it
alone until $\Gamma _i < 0.01\Gamma$. Negative distances, and
orbits that cross at zero separation, are unphysical but allowed
by the mathematics. During the relaxation a distance can move
from negative to positive, so when object $i$ has distance 
$d_i< 30$~kpc and $\Gamma _i <1$ we choose a new random orbit for
this object and relax it until $\Gamma _i < 0.01\Gamma$. 
When $\Gamma <1$ and the minimum
comoving separation of a pair of particles is less than 30~kpc we
apply the above procedure to the less massive one.
Equation~(\ref{eq:dddz}) assumes the coordinates are at a
stationary point of $A$, but we 
apply the redshift adjustment at each coordinate adjustment
starting from random orbits. When the coordinates are far
from a stationary point the approximation to 
$\p ^2\chi ^2/\p z^2$ in equation~(\ref{eq:chipp}) can be
negative. In this case we relax 
down the gradient of $\chi ^2$, using
\beq
\delta z = -\epsilon {\chi ^2\over \p\chi ^2/\p z},
\label{eq:deltazg}
\eeq
When the second derivative of $\chi ^2$ is positive
we use the smaller of the redshift adjustments from
equations~(\ref{eq:deltaz}) and~(\ref{eq:deltazg}). There
are many solutions to the equation of motion for the objects
in Tables~1 and ~2; we seek those with redshifts and distances
that are close to the catalog values. When the relaxation brings
$\Gamma$ below $10^{-5}$ each object in turn with individual
$\chi ^2$ greater than 25 is given 40 attempts at a new orbit,
relaxed from a random one, to bring the individual $\chi
^2$ below 25. After this treatment of each offending orbit
all orbits are iteratively relaxed until $\Gamma <10^{-5}$
again. The operation is repeated, allowing a maximum of 5 sets
of 40 attempts for each object.  In the analysis of the real
catalogs usually all individual $\chi ^2$ are below 25 after
about 1000 iterations through all orbits. In the control case
with randomly placed present angular positions the allowed
number of attempts usually is exhausted by 1000 iterations,
leaving some objects with quite large $\chi ^2$. At $n_x>10$
time steps the computation time scales as $n_x^3$, dominated by
the matrix inversion codes LUDCMP and LUBKSB from \cite{Press}
(1992). 

\section{Application to the Nearby Galaxies}

\subsection{Parameters and Control Samples}

In this illustration of the numerical method we fix some
parameters by considerations 
other than the motions of the nearby galaxies. We adopt a
cosmologically flat Friedmann-Lema\^\i tre model with matter
density and Hubble parameters
\beq
\Omega_m = 0.25,\qquad 
H_{\circ}=75h_{75}\hbox{ km s}^{-1}~\hbox{Mpc}^{-1}.
\label{eq:cos_par}
\eeq
Our choice of $\Omega_m$ is near the lower end of the range
derived from recent measurements of the angular fluctuations
in the thermal cosmic background radiation (eg. \cite{Hu}
2000) and near the upper end from analyses of galaxy dynamics (as
summarized by \cite{TS98} 1998 and \cite{Bahcall} 2000). The value
of $H_{\circ}$ fits the recession velocities of galaxies at 
$cz=5000$ to 8000 km~s$^{-1}$ that plausibly are in the Hubble
flow (\cite{TP00} 2000). The scale factor $h_{75}$ takes
account of a close to common uncertainty in the distance zero
points for $H_o$ and the distances in Tables~1 and~2.

Luminosities are computed at the catalog distances. The masses in
the inner zone in Table~1 assume the mass-to-light ratio 
\beq
M/L({\rm inner})=75h_{75}~M_\odot /L_\odot .
\label{eq:MoLi}
\eeq
This choice makes the sum of the masses of M31 and the Milky Way
consistent with the spherical model for their relative motion,
and the total mass contrast at
$r<4h_{75}^{-1}$~Mpc is $\delta M/\langle M\rangle =1.0$,
consistent with the modest gravitational slowing of the local
expansion rate just 
outside the Local Group. Following the discussion in \S~2, we
take account of the light in the outer zone that is not in the
mass concentrations in Table~2 by adopting a larger mass-to-light 
ratio, 
\beq
M/L({\rm outer})=150h_{75}~M_\odot /L_\odot ,
\label{eq:MoLo}
\eeq
for the late-type systems in Table~2, and 
$750h_{75}~M_\odot /L_\odot$ for the early type systems. This
makes the contrast  
$\delta M/\langle M\rangle =0.6$ at $r<20h_{75}^{-1}$~Mpc, again
a reasonable number. 

To take account of the distributed mass around the tracers we 
change the first term in the gravitational potential in
equation~(\ref{eq:V}) to 
$V\propto (x_{ij}{}^2 +(c_i + c_j)^2)^{-1/2}$, with  
comoving cutoff length $c_i=30$~kpc for all particles in the
inner zone, 2~Mpc for Virgo, and 700~kpc for all other objects in
the outer zone. The cutoff is larger in the outer zone because
these objects are meant to represent more broadly distributed
mass. When all cutoff lengths are set to zero it allows more
sharply curved and perhaps unphysical orbits, but the statistics
are little changed. 

The standard deviations $\sigma _{\rm cz}$ in redshift
take account of the dispersions of velocities in the large
systems of galaxies in Table~2, but in Table~1 reflect only
the measurement errors. Since we can only guess at the typical
difference between the velocity of the galaxy and the motion of
the center of mass of the dark halo it is supposed to represent,
we write the standard deviation in redshift as
\beq
\sigma _z=(\sigma _{cz}^2 + \sigma _o^2)^{1/2},
\label{eq:sigmaz}
\eeq
and we compare results for $\sigma _o=0$ and $c\sigma _o=15$
km~s$^{-1}$.

We label dynamical solutions for the combined set of orbits of
the objects in both zones as c0 and c15, for the two values of
$\sigma _o$. In the first set of control cases, labeled i0 and i15, 
the outer zone has been removed from the dynamics. 
In the second set of control cases, labeled r0 
and r15, the outer zone is included and the catalog distances
and redshifts are used but each object is placed at a randomly chosen
position in the sky (with the same angular position in all
solutions). All statistics for all cases are
based on the objects in the inner zone alone.

We use $n_x=20$ time steps uniformly spaced in the 
expansion parameter $a(t)$ and $\epsilon = 0.2$ in the
relaxation equations~(\ref{eq:deltax}), (\ref{eq:deltaz}),
and~(\ref{eq:deltazg}). The statistics are quite similar
at $\epsilon = 0.1$ and $\epsilon = 0.3$, and at $n_x=15$ and
$n_x=30$. We compute 30 solutions from random orbits for each
case.   

\begin{figure}[ht]
\plotone{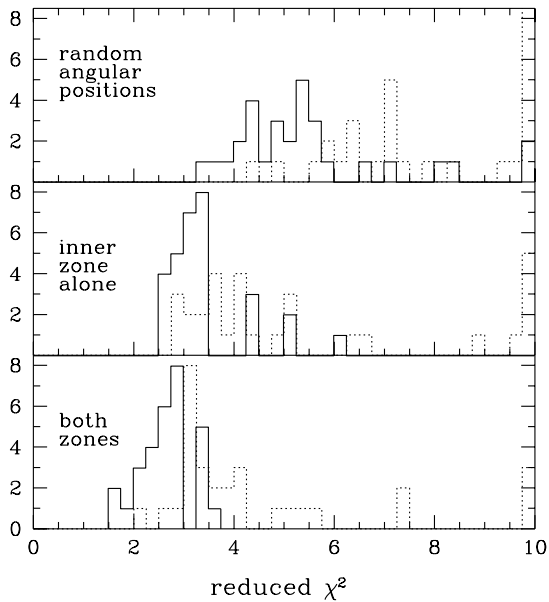}
\caption{Distributions of reduced $\chi ^2$ in distance and
redshift for 30 solutions for each of 
the six cases. In the broken histograms 
$\sigma _o=0$, in the solid, $c\sigma _o=15$ km~s$^{-1}$.}
\end{figure}

\subsection{Numerical Solutions}

Figure~2 shows the distributions of reduced $\chi ^2$
in redshift and distance summed over the 29 objects in the
inner zone (excluding the Milky Way) for the 30 solutions
for each of the six cases. Increasing the standard deviations in
the 
redshifts makes it considerably easier to find solutions with 
reduced $\chi ^2$ less than 3.  Removing the outer zone
from the dynamics or scrambling the angular positions
makes it more difficult to find these relatively good fits to
the data. The differences of the best $\chi ^2$ among the 30 
solutions for each case are much smaller than the
differences among the distributions of $\chi ^2$ from all 30
solutions, but the best solution for the real catalogs is
consistently better than the best solution for the control cases. 

\begin{figure}[ht]
\plotone{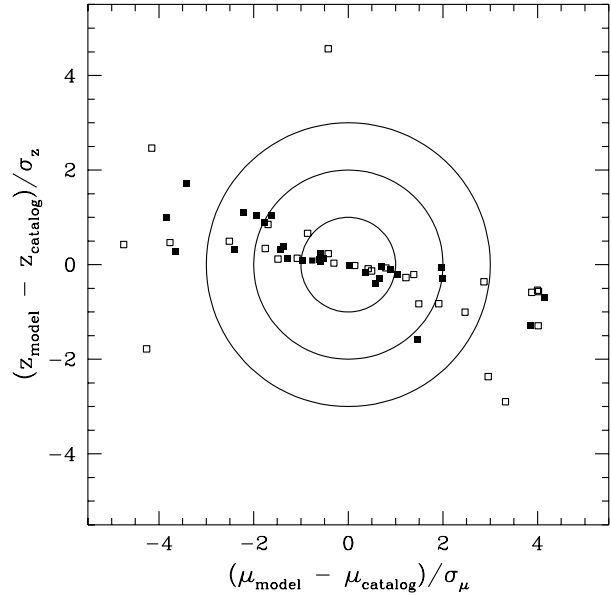}
\caption{Distributions of normalized differences between model
and catalog redshifts and distance moduli in the best solutions
for cases c0 (filled squares) and r0 (open squares).}
\end{figure}

The last two columns in Table~1 and the filled squares in
Figure~3 show the normalized differences between model and
catalog redshifts and distances for the 29 objects in the
best solution for the real catalogs with $\sigma _o =0$ (case
c0). For comparison we show as the open symbols
the normalized differences in the best of the solutions for
scrambled angular positions (case r0). The normalized scatter is
smaller in redshift and correlated with the scatter in
distance. This is because $\p d/\p z$ (eq.~[\ref{eq:dddz}]),
the rate of change of redshift with respect to distance in a
family of solutions, usually is positive and, in the units in
Figure~3, the slope usually is greater than unity. The single-particle 
$\chi ^2$ are minimized at the closest approach of the trajectory
of redshift as a function of distance to the origin in Figure~3,
so the minima scatter around a line that trends down 
to the right. Some trajectories happen to pass close to the
origin in Figure~3; their 
relaxation to minimize $\chi ^2$ produces unrealistically
good fits to the catalog redshifts and distances.    

\begin{figure}[ht]
\plotone{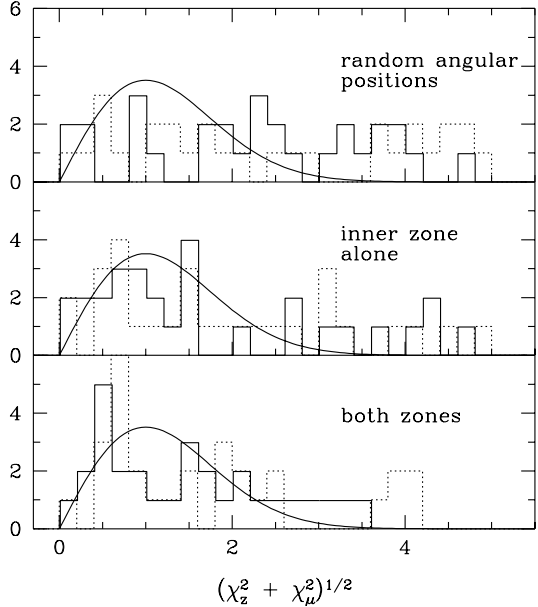}
\caption{Distributions of single-particle $\chi ^2$ in redshift
and distance modulus for the 29 objects in the solutions with
the smallest total $\chi ^2$ for each case. In the broken histograms 
$\sigma _o=0$, in the solid, $\sigma _o=15$ km~s$^{-1}$.}
\end{figure}

Figure~4 shows the distributions of single-particle $\chi ^2$
in redshift and distance for the best solutions for each of the
six cases. (To avoid confusion we note that Fig.~2
is the distribution of total $\chi^2$ among 30 solutions, while
Fig.~4 is the distribution of single-particle $\chi ^2$
for the 29 objects in the best solution for each case.) To reduce
the spread of values we plot the square 
root of the sum of the squares of the normalized differences
of model and catalog redshifts and distances. The curve is the
distribution for Gaussian normal scatter of the redshifts and
distances. There tend to be more objects with $\chi ^2 <1$
than expected for this idealized model, even in the random case.
This is a result of the relaxation to minimize $\chi ^2$, as
discussed in connection with Figure~3. Even in the best solution
for the real data there are more large values of $\chi ^2$ than
expected 
from the idealized distribution, but the excess is not large. For  
example, in c0 there are 10 of the 29 objects with 
$\chi ^2>4$, about 2.5 times the number expected from the
idealized model. 

Our computation allows many attempts to find an acceptable
fit to catalog redshifts and distances, and the fit may be
achieved at the expense of an unrealistic initial condition, 
associated with an unacceptable value or gradient
of the primeval mass density fluctuation. Estimates of the
primeval density fluctuations associated with a numerical
action solution are discussed in \cite{Peeb96} (1996). We
use a simplified approach based on the linear perturbation
relation between the mass density contrast $\delta$ and the
peculiar velocity field $\vv (\xv )$, \beq {\p\delta\over\p t} =
-{1\over a}\nabla\cdot\vv .  \eeq This expression averaged over
a sphere of comoving radius $x$ is $\dot\delta = -3v_r/ax$,
where $v_r$ is the peculiar velocity normal to and averaged
over the surface of the sphere. The density contrast at high
redshift varies as 
$\delta\propto D(t)$, where $D(t)$ is the growing solution
to the linear perturbation equation, so an estimate of the
local density contrast extrapolated to the present in linear
perturbation theory is
\beq
\delta _o = -3{D_o\over\dot D(t)}{(\dot\xv _i-\dot\xv _j)\cdot
(\xv _i-\xv _j)\over x_{ij}^2}.
\label{eq:deltao}
\eeq
The dot means derivative with respect to proper world time
$t$, $D_o$ is the present value of $D$, and $\xv _i$ and
$\dot\xv _i$ are the coordinate position and velocity of
particle $i$ at time $t$. We use the approximation 
$\dot\xv = (\xv _3-\xv _2)/(t_3 - t_2)$, from the second to the
third time step, because the fractional increase in time from the 
first to the second step is large, though it yields similar
results.  

\begin{figure}[ht]
\plotone{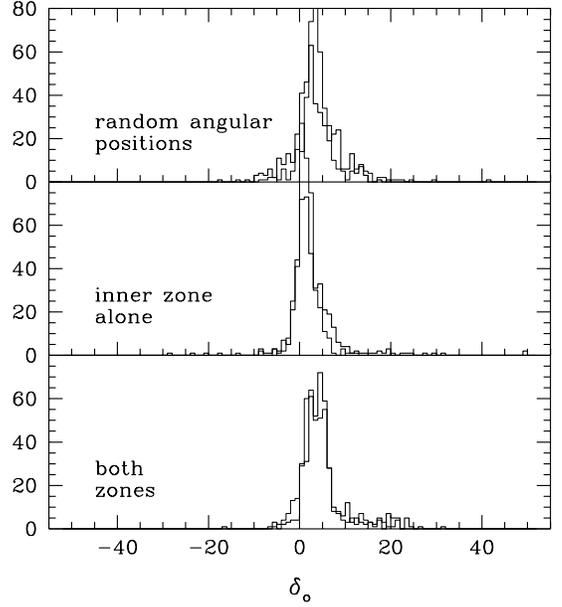}
\caption{Distributions of estimates of the local primeval density
contrast extrapolated to the present in linear perturbation
theory (eq.~[\ref{eq:deltao}]) in the best solutions for the
six cases. The thin histograms assume $\sigma _o=0$,
the thick, $c\sigma _o=15$ km~s$^{-1}$.} 
\end{figure}

Figure~5 shows distributions of $\delta_o$ from all pairs of
objects in the inner catalog for the best solution 
for each of the six cases. When the outer zone is removed from
the dynamics the central value of the contrast $\delta _o$
is close to unity, the value expected from the choice 
of $M/L$. The outer zone objects increase the central value of
$\delta _o$ for the inner zone, we suspect because the inner and
outer zone objects are somewhat intermingled at high redshift.
There are substantial differences
among the distributions from solutions with similar values
of $\chi ^2$; reproducible features include the smaller central
value of $\delta _o$ when the outer zone is removed, the
prominent positive tail in the distribution of $\delta _o$ from
the real catalogs, and the still more prominent tails
in the random case. 

Because we have used a crude estimate of $\delta _o$, and
lumped together a range of values of primeval 
separations $x_{ij}$, we could not expect the distributions in
Figure~5 to be Gaussian even if we had an adequate approximation
to Gaussian initial conditions. However, we might expect the more
realistic cases to be closer to Gaussian. By this criterion the 
random control case is the least realistic, as we would hope. On
the other hand, the outer zone increases the positive tail for 
the objects in the inner zone. This illustrates the difficulty of
testing models at the present level of accuracy of the redshift
and distance measurements. We turn now to a more
demanding test. 

\begin{figure}[ht]
\plotone{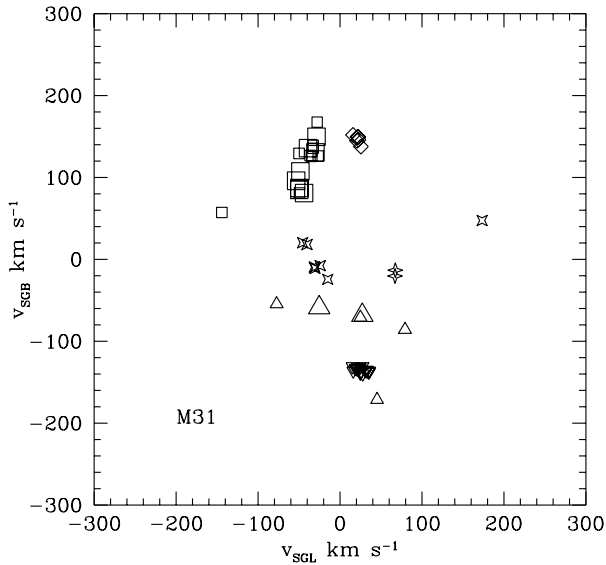}
\caption{Transverse galactocentric velocity of M31 in all 30
solutions. The larger symbols have reduced total $\chi^2$ less
than 2.5. Squares, crosses and triangles pointing up use
the two zones, case c15. Diamonds, plus signs,
and triangles pointing down show the effect of removing the outer 
zone. The symbol shapes are chosen according to $v_{\rm SGB}$,
to facilitate comparison with the next 2 figures.}
\end{figure}

\section{Proper Motions}

The satellite missions SIM (http://sim.jpl.nasa.gov) and GAIA 
(http://astro.estec.esa.nl/GAIA)
could measure the velocities of nearby
galaxies relative to the Milky Way and perpendicular to the line
of sight, and thereby test the assumption that the galaxies trace
the local mass distribution. 

Figures 6 to 8 show examples of transverse velocities in our 30
solutions The larger symbols represent the better fits to the
catalog redshifts and distances, with reduced  $\chi ^2$ for the
inner objects less then 2.5, the smaller symbols the solutions
with larger  $\chi ^2$. The symbol orientations distinguish
solutions with and without the outer zone objects in the
dynamics, as explained in the caption to Figure~6. 

When the dynamics includes only the inner catalog the transverse
galactocentric velocity of M31 (Fig.~6) is close to zero in a
few solutions, but in most cases M31 is moving at close to  
$\pm 150$ km~s$^{-1}$ normal to the supergalactic plane. This is 
part of a collapse to the local sheet-like arrangement shown in 
Figure~1. When the outer zone objects are included in the
dynamics the fit is improved, the solutions have a broader range
of values of the transverse velocity, and there are more solutions
with a near radial approach of M31. The better solutions still
prefer motion toward the supergalactic plane, and there
are many possible directions of transverse motion that could
appear to falsify the action model. 

\begin{figure}[ht]
\plotone{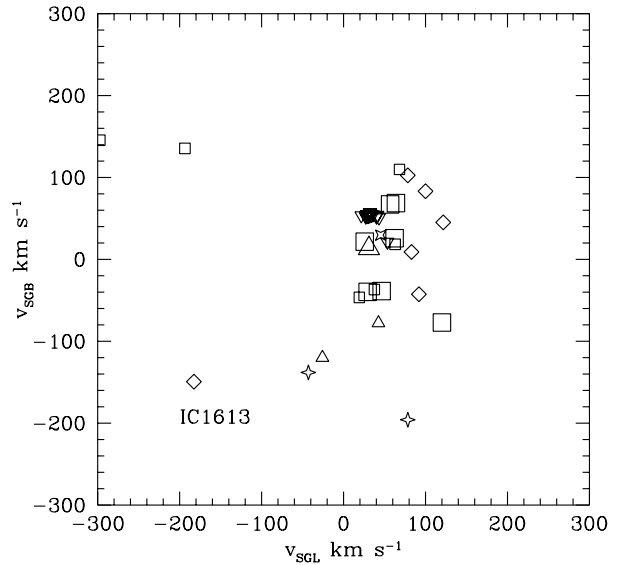}
\caption{Transverse galactocentric velocity of
IC~1613. The symbol shapes are chosen according to the value of
${\rm v_{SGB}}$ for M31, and the sizes according to the value of
the total $\chi ^2$, as in Fig.~6.} 
\end{figure}

The distribution of solutions for the transverse galactocentric
velocity of IC~1613 in Figure~7 is broader when
the outer zone is included, as for M31. One might have expected
to see a correlation of solutions for the velocities of M31 and
this distant companion of M31. There is no pronounced correlation
in this example, but we regard this as a very preliminary
indication to be considered in more detail. 

\begin{figure}[ht]
\plotone{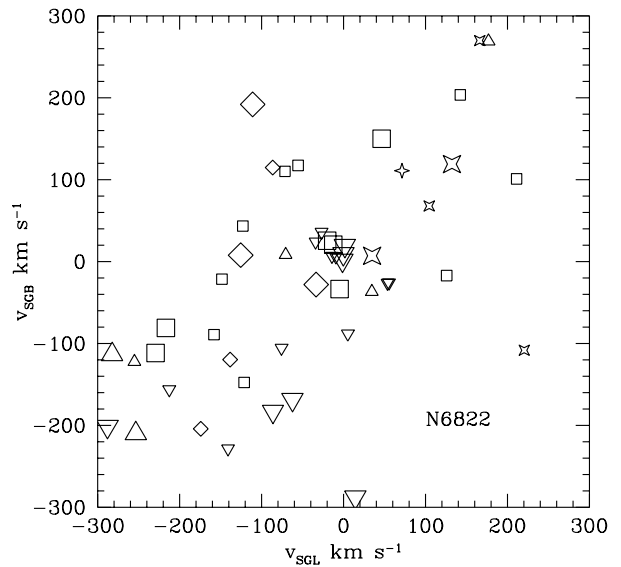}
\caption{Transverse galactocentric velocity of
NGC~6822. The symbol shapes are chosen according to the value of
${\rm v_{SGB}}$ for M31 in Fig.~6, but here the larger symbols
represent solutions in which the radial velocity of NGC~6822 
is within 30 km~s$^{-1}$ of the catalog value.} 
\end{figure}

The orbit of NGC~6822 is more complex than
M31 because NGC~6822 is closer and moving away. 
Figure~8 shows one can find acceptable fits
to the catalog redshift and distance, with a broader
range of possible proper motions than in the previous two
examples. We suspect this illustrates a loss of memory of
initial conditions in nonlinear classical dynamics. Tighter
constraints on acceptable solutions might arise from a study of
the initial conditions in the solutions; this also requires more
study. 

\section{Discussion}

\subsection{The Assumption Galaxies Trace Mass}

If the nearby galaxies trace the mass on scales $\gsim 100$~kpc
then our inner catalog in Table~1 gives a reasonably complete 
description of the mass within 4$h_{75}^{-1}$~Mpc,
and Table~2 gives a first approximation to the effect
of external masses on the Local Group and its immediate
neighbors. We have found dynamical solutions for the orbits 
of the assumed mass tracers that end up in our inner zone (within
4$h_{75}^{-1}$~Mpc distance); in the best cases the rms
difference between model and catalog redshifts and distances is
in the range 1.25 to 1.5 
times the nominal catalog errors. The better fits use a more
liberal estimate for the redshift errors, an issue that requires
further consideration. The difference of 
the reduced $\chi ^2$ from unity is considerably larger than
statistical, but small enough to argue for the assumption
galaxies trace mass.  

One of our controls keeps the catalog distances,
redshifts, and masses but places the objects at random positions
in the sky. Since this preserves the near Hubble flow of the
more distant members of the inner zone and the infall of
most of the closer members, it is not surprising that we get
a rough fit. But this scrambling of 
the present angular positions worsens the fit, as expected
if galaxies trace mass. This result is quite stable under changes 
in the random numbers for the scrambling and computation.

Our second control removes the outer masses from the dynamics. It
shows that the outer masses improve the fit to the
measured redshifts and distances in the inner zone, again
as expected if galaxies trace 
mass. This also suggests our solutions for the inner zone
would be improved by using a better model for the external
mass distribution, another point that requires further
consideration. 

Figures 6 to 8 show that, if galaxies trace mass, there are
solutions with similar values of $\chi ^2$ and transverse
velocities that differ by 300 km~s$^{-1}$. These solutions tend
to occupy restricted regions of transverse velocity space.
The predictions for the galactocentric transverse velocity of M31
are only weakly perturbed by the masses beyond
4$h_{75}^{-1}$~Mpc. Thus we suspect that under the assumption
galaxies trace mass the allowed velocity of M31 is rather tightly 
constrained by the present catalogs. A measurement of the 
transverse velocity of this and other nearby galaxies could
critically test the picture of the local mass distribution.  

\subsection{Comments on Future Work}

This new version of the numerical action method could and
certainly should be applied to numerical N-body simulations. 
We hope to continue work on this topic. Our exploration of the
effect of adjusting the mass-to-light ratios in the data has been
quite limited, and here too we hope to be able to report on the
values of $M/L$ for late and early type systems that yield the
best fit to catalog redshifts and distances, to compare to global
measures of the density parameter.   

The allowance for errors in both redshift and distance is more
realistic than previous applications of the action method, 
but it still ignores the errors in masses. A study of this
issue certainly will be part of a satisfactory  assessment of the
dynamics of the nearby galaxies. 

A fundamental limitation of our numerical application should be
noted. The matrix inverse of the second derivatives of the action
with respect to the coordinates of an orbit gives the rate of
change of the present distance with respect to the redshift
(eq.~[\ref{eq:dddz}]), and it reaches saddle points and
local maxima as well as the local minima that are reached by
following the gradient of the action. Because 
the computation time for the matrix inverse varies as the cube
of the matrix size we relax the coordinates for each
particle separately, toward a stationary point of the action and
a minimum of the single-particle $\chi ^2$. This is iterated
through all particles many times. We are really interested in the
minimum of the sum of the $\chi ^2$ for all particles, which need
not minimize any single-particle $\chi ^2$. The inversion of the
much larger matrix of all coordinates of all particles would be a 
heavy computation, but worth attempting.   

The galaxy NGC~6822 illustrates the problem of dealing with
complex orbits. We have found reasonably close fits to its
positive radial velocity by relaxing from random orbits, but
it would be feasible and useful to develop a more systematic
sample of the many allowed orbits for such cases. The results
would challenge us to decide which solutions could have come from  
realistic initial conditions; we need something better than the
simple estimator in equation~(\ref{eq:deltao}).  

\acknowledgements

This work was
supported in part at Princeton University by the NSF.

\onecolumn

\begin{center}
\begin{tabular*}{1.1\textwidth}%
{lcrrcrrrcrr}
\multicolumn{11}{c}{Table 1. Inner Catalog}\\
\tableline
Name &Note & SGL & SGB  & distance & $\sigma _\mu$\ & redshift  & $\sigma _{cz}$ 
& mass & $\chi _\mu$ & $\chi _z$ \\
& &(deg) &(deg) & (Mpc) & & (km s$^{-1})$ & (km s$^{-1})$ & ($\log_{10}[M_\odot$])  \\
\tableline
I342    & +a & $ 10.8$ & $  0.3$ & $  3.6$ & $0.59$ & $ 171$ & $ 5$ &
$12.5$ & $ 2.0$ & $-0.1$ \cr 
Maffei  & +b & $359.4$ & $  1.2$ & $  4.0$ & $0.40$ & $ 152$ & $54$ &
$12.5$ & $ 0.6$ & $-0.4$ \cr 
M31     & +c & $335.5$ & $ 11.1$ & $  0.8$ & $0.16$ & $-113$ & $ 5$ &
$12.5$ & $-1.6$ & $ 1.0$ \cr 
M81     & +d & $ 41.1$ & $  0.7$ & $  3.7$ & $0.17$ & $ 114$ & $ 6$ &
$12.4$ & $-1.4$ & $ 0.4$ \cr 
N5128   & +e & $161.6$ & $ -7.0$ & $  3.5$ & $0.31$ & $ 371$ & $ 9$ &
$12.4$ & $ 1.0$ & $-0.2$ \cr 
GALAXY  &    & $ - $&$ - $&$ - $&$ - $&$ - $&$ - $&                  
$12.3$&$ - $&$ - $ \cr 
N253    & +f & $272.6$ & $ -4.7$ & $  2.9$ & $0.30$ & $ 229$ & $ 5$ &
$12.2$ & $-0.8$ & $ 0.1$ \cr 
N5236   & +g & $148.6$ & $  0.4$ & $  3.3$ & $0.56$ & $ 373$ & $ 8$ &
$11.9$ & $ 0.9$ & $-0.1$ \cr 
N2403   & +h & $ 30.8$ & $ -7.5$ & $  3.3$ & $0.20$ & $ 227$ & $ 5$ &
$11.8$ & $ 4.1$ & $-0.7$ \cr 
N4236   &    & $ 47.1$ & $ 11.4$ & $  3.9$ & $0.30$ & $ 118$ & $ 9$ &
$11.5$ & $-3.8$ & $ 1.0$ \cr 
Circinus&    & $183.1$ & $ -6.4$ & $  3.3$ & $0.80$ & $ 268$ & $ 9$ &
$11.5$ & $-0.6$ & $ 0.1$ \cr 
N55     & +i & $256.2$ & $ -2.4$ & $  1.6$ & $0.30$ & $  96$ & $ 5$ &
$11.5$ & $-0.5$ & $ 0.1$ \cr 
N7793   &    & $261.3$ & $  3.1$ & $  4.1$ & $0.30$ & $ 230$ & $ 7$ &
$11.5$ & $-2.4$ & $ 0.3$ \cr 
N300    &    & $259.8$ & $ -9.5$ & $  2.1$ & $0.20$ & $ 101$ & $ 5$ &
$11.3$ & $-3.4$ & $ 1.7$ \cr 
N1569   &    & $ 11.9$ & $ -4.9$ & $  2.4$ & $0.60$ & $  47$ & $ 5$ &
$11.0$ & $ 0.7$ & $ 0.0$ \cr 
N404    &    & $331.8$ & $  6.3$ & $  3.6$ & $0.20$ & $ 115$ & $13$ &
$10.8$ & $-0.6$ & $ 0.2$ \cr 
N3109   & +j & $138.0$ & $-45.1$ & $  1.3$ & $0.17$ & $ 194$ & $ 5$ &
$10.5$ & $-1.4$ & $ 0.3$ \cr 
IC5152  &    & $234.2$ & $ 11.5$ & $  1.7$ & $0.20$ & $  85$ & $ 9$ &
$10.0$ & $-1.9$ & $ 1.0$ \cr 
SexA,B  & +k & $102.6$ & $-40.2$ & $  1.4$ & $0.17$ & $ 165$ & $ 6$ &
$10.0$ & $ 3.8$ & $-1.3$ \cr 
N6822   &    & $229.1$ & $ 57.1$ & $  0.5$ & $0.17$ & $  45$ & $ 6$ &
$10.0$ & $-1.8$ & $ 0.9$ \cr 
N1311   &    & $243.2$ & $-34.3$ & $  3.1$ & $0.40$ & $ 425$ & $15$ &
$ 9.8$ & $ 2.0$ & $-0.3$ \cr 
IC1613  &    & $299.2$ & $ -1.8$ & $  0.7$ & $0.17$ & $-155$ & $ 5$ &
$ 9.8$ & $-3.7$ & $ 0.3$ \cr 
WLM     &    & $277.8$ & $  8.1$ & $  0.9$ & $0.17$ & $ -65$ & $ 5$ &
$ 9.7$ & $-2.2$ & $ 1.1$ \cr 
VIIZw403&    & $ 36.9$ & $ 11.4$ & $  4.5$ & $0.20$ & $  50$ & $ 7$ &
$ 9.6$ & $-0.6$ & $ 0.1$ \cr 
\tableline
\end{tabular*}
\end{center}
\begin{center}
\begin{tabular*}{1.1\textwidth}%
{lcrrcrrrcrr}
\multicolumn{11}{c}{Table 1. (Continued) Inner Catalog}\\
\tableline
Name &Note & SGL & SGB  & distance & $\sigma _\mu$\ & redshift  & $\sigma _{cz}$ 
& mass & $\chi _\mu$ & $\chi _z$ \\
& &(deg) &(deg) & (Mpc) & & (km s$^{-1})$ & (km s$^{-1})$ & ($\log_{10}[M_\odot$])  \\
\tableline
SagDIG  &    & $221.3$ & $ 55.5$ & $  1.1$ & $0.20$ & $   7$ & $ 7$ &
$ 9.1$ & $ 0.4$ & $-0.2$ \cr 
PegDIG  &    & $305.8$ & $ 24.3$ & $  0.8$ & $0.20$ & $ -22$ & $ 5$ &
$ 8.9$ & $-1.0$ & $ 0.1$ \cr 
LeoA    &    & $ 69.9$ & $-25.8$ & $  0.7$ & $0.30$ & $ -18$ & $ 7$ &
$ 8.8$ & $ 0.7$ & $-0.3$ \cr 
DDO210  &    & $252.1$ & $ 50.2$ & $  0.9$ & $0.20$ & $ -24$ & $ 6$ &
$ 8.7$ & $ 0.0$ & $ 0.0$ \cr 
DDO155  &    & $103.0$ & $  4.7$ & $  1.5$ & $0.40$ & $ 183$ & $ 5$ &
$ 8.6$ & $-1.3$ & $ 0.1$ \cr 
Phoenix &    & $254.3$ & $-20.9$ & $  0.4$ & $0.20$ & $ -34$ & $10$ &
$ 7.7$ & $ 1.5$ & $-1.6$ \cr 
\tableline
\end{tabular*}
\end{center}
Notes to Table 1. The following galaxies are combined into single entries.

a) IC 342, NGC 1560, UGCA 105

b) Maffei 1, Maffei 2

c) M 31, M 33, IC 10, LGS 3

d) M 81, M 82, NGC 2976, IC 2574

e) NGC 5128 = Cen A, NGC 4945

f) NGC 253, NGC 247

g) NGC 5236 = M 83, NGC 5102 

h) NGC 2403, NGC 2366, Holmberg II

i) NGC 55, ESO 294-010

j) NGC 3109, Antlia dwarf

k) Sextans A, Sextans B

\begin{center}
\begin{tabular*}{1.1\textwidth}%
	{lcrrrrrrcc}
\multicolumn{10}{c}{Table 2. Outer Catalog}\\
\tableline
Name & Group & SGL & SGB  & distance & $\sigma _\mu$ & redshift & $\sigma _{cz}\ \ $ 
& mass & mass/dist$^3$ \\
& &(deg)&(deg)& (Mpc) & & (km/s) & (km/s) & ($\log_{10}[M_\odot$]) &
$(10^{10}M_\odot$ Mpc$^{-3}$) \\
\tableline
Virgo     & 11~-1 & $102.7$ & $ -2.4$ & $ 17.0$ & $0.17$ & $ 986$ & $63$ &
$15.1$ & $24$ \cr 
Fornax    & 51~-1 & $262.2$ & $-40.9$ & $ 18.8$ & $0.20$ & $1466$ & $78$ &
$14.5$ & $ 4$ \cr 
Dorado    & 53~-1 & $234.6$ & $-40.4$ & $ 18.5$ & $0.20$ & $1076$ & $65$ &
$14.3$ & $ 3$ \cr 
Coma I    & 14~-1 & $ 87.0$ & $  1.2$ & $ 16.4$ & $0.40$ & $ 983$ & $53$ &
$14.1$ & $ 3$ \cr 
Ursa Major& 12~-1 & $ 65.8$ & $  2.6$ & $ 16.6$ & $0.20$ & $ 957$ & $20$ &
$13.9$ & $ 2$ \cr 
Leo       & 15~-1 & $ 93.7$ & $-25.8$ & $ 11.1$ & $0.17$ & $ 722$ & $37$ &
$13.8$ & $ 4$ \cr 
Sombrero  & 11-14 & $126.2$ & $ -6.7$ & $ 10.0$ & $0.30$ & $ 985$ & $47$ &
$13.5$ & $ 3$ \cr 
N6744     & 19~-1 & $207.8$ & $ 10.2$ & $ 13.9$ & $0.40$ & $ 752$ & $31$ &
$13.3$ & $ 1$ \cr 
CVn II    & 14~-4 & $ 70.5$ & $  5.6$ & $  7.7$ & $0.20$ & $ 556$ & $12$ &
$13.0$ & $ 2$ \cr 
M51       & 14~-5 & $ 73.0$ & $ 16.1$ & $  8.0$ & $0.40$ & $ 562$ & $46$ &
$13.0$ & $ 2$ \cr 
N1023     & 17~-1 & $340.4$ & $ -8.0$ & $ 10.0$ & $0.20$ & $ 696$ & $10$ &
$13.0$ & $ 1$ \cr 
M101      & 14~-9 & $ 63.3$ & $ 22.6$ & $  7.4$ & $0.20$ & $ 354$ & $27$ &
$12.9$ & $ 2$ \cr 
N6946     & 14~-0 & $ 10.0$ & $ 42.0$ & $  6.0$ & $0.60$ & $ 271$ & $ 6$ &
$12.8$ & $ 3$ \cr 
CVn I     & 14~-7 & $ 76.4$ & $  6.1$ & $  4.9$ & $0.30$ & $ 316$ & $11$ &
$12.6$ & $ 4$ \cr 
\tableline
\end{tabular*}
\end{center}

\end{document}